\documentclass[aps,prl,amsmath,amsfonts,amssymb,twocolumn]{revtex4}

\usepackage[latin1]{inputenc}
\usepackage{graphicx}
\usepackage[english]{babel}
\usepackage[usenames,dvipsnames]{color}
\usepackage{amsfonts}
\usepackage{amsmath}
\usepackage{amssymb}

\newcommand{\beq}{\begin{equation}}
\newcommand{\eeq}{\end{equation}}
\newcommand{\nn}{\nonumber}
\newcommand{\ket}[1]{|#1\rangle}
\newcommand{\bra}[1]{\langle #1|}

\newcommand{\ra}{\rightarrow}

\makeatletter
\@ifundefined{textcolor}{}
{%
 \definecolor{BLACK}{gray}{0}
 \definecolor{WHITE}{gray}{1}
 \definecolor{RED}{rgb}{1,0,0}
 \definecolor{GREEN}{rgb}{0,1,0}
 \definecolor{BLUE}{rgb}{0,0,1}
 \definecolor{CYAN}{cmyk}{1,0,0,0}
 \definecolor{MAGENTA}{cmyk}{0,1,0,0}
 \definecolor{YELLOW}{cmyk}{0,0,1,0}
 }

\begin{document}


\title{Self-replication of a quantum artificial organism driven by single-photon pulses}

\author{Daniel Valente
$^{1}$
}
\email{valente.daniel@gmail.com}



\affiliation{
$^{1}$ 
Instituto de F\'isica, Universidade Federal de Mato Grosso, Cuiab\'a 78060-900 Mato Grosso, Brazil
}


\begin{abstract}
Artificial organisms are computer programs that self-replicate, mutate, compete and evolve.
How do these lifelike information-processing behaviours could arise in diverse far-from-equilibrium physical systems remains an open question.
Here, I devise a toy model where the onset of self-replication of a quantum artificial organism (a chain of lambda systems) is owing to single-photon pulses added to a zero-temperature environment.
The model results in a replication probability that is proportional to the absorbed work from the photon, in agreement with the theory of dissipative adaptation.
Unexpectedly, spontaneous mutations are unavoidable in this model, due to rare but finite absorption of off-resonant photons.
These results hint at self-replication as a possible link between dissipative adaptation and open-ended evolution.
\end{abstract}

\maketitle

\section{Introduction}
The longstanding question concerning the general principles of life still inspires a large diversity of concepts, methods and viewpoints \cite{schrodinger,bridge1, cleland, englandbook}.
Besides its fundamental value, for instance towards a universal biology \cite{nigel}, looking for generalities could eventually enable us to imitate and extend the sophisticated dynamics achieved by living things.
The artificial life approach, for instance, circumvents the biochemical constraints of living organisms by studying computer programs that mimic distinctive features of life.
Artificial organisms can harness self-replication so as to mutate, compete and evolve complex features \cite{lenski,OEE19}.
When one is interested in the most essential aspects of artificial life, quantum mechanics can be instrumental in casting the computation as operations on a handful of quantum bits \cite{unai1,unai2,unai3,lamata}.
Still, the finely engineered character of the computers assumed from the outset leaves open the question of what principles could explain the self-assembly of lifelike information-processing systems from physical laws.

The nonequilibrium thermodynamic approach, by contrast, looks for the emergence of lifelike behaviours in driven physical systems, comprising disordered, self-assembling, and self-replicating ones \cite{JCP2013, naturenano2015,englandbook, science2010, scirep2013, PRE2015, natmat2017, naturephoton2018, englandPRE2019, naturenano2020}.
Remarkably, it has been recently proposed that nonequilibrium thermodynamics could generalize Darwinian evolution, even for non-replicating systems \cite{prx2016}. 
The idea is that exceptional specialization, or fine-tuned adaptation, to an environment by a fluctuating physical system can be fueled by the irreversible work consumption along far-from-equilibrium trajectories \cite{PNAS2017,PRL2017}.
This exceptional self-organized response of a driven system to the patterns of its environment has been called dissipative adaptation \cite{naturenano2015,englandbook,naturenano2020}.
Quantum physics can, once more, help us to showcase the most elementary aspects of dissipative adaptation \cite{dv21}.
Yet, thermodynamic studies of adaptation have been focusing mostly on systems that stabilize in the out-of-equilibrium states, lacking a more thorough discussion on the origins of diversification \cite{natchem} and open-ended evolution \cite{OEE,OEE19} akin to that achieved with artificial life, and necessary for explaining the ever-growing complexity and diversity of biological species.

Can a driven physical system undergoing dissipative adaptation implement an artificial-life computational operation?
This would provide a further step in the connections between the information-processing underlying self-replication and the dissipation of energy inherent to metabolism \cite{copol,copolj,ti}, ultimately helping us to imitate the transition from inanimate to living matter \cite{bridge1,algorithmicorigins} in diverse physical systems.
More broadly, searching for a lifelike self-organized computation could also contribute to the recently proposed goal of `thermodynamic computing' \cite{tc}.

Here, I formulate a toy model inspired by this idea of an artificial-life code with a driven system undergoing dissipative adaptation.
Namely, the self-replication of a minimal quantum artificial organism starts due to single-photon pulses added to a zero-temperature environment.
The model results in a replication probability that is linearly proportional to the average work absorbed by the replicating organism, characterizing a quantum dissipative adaptation.
Counterintuitively, spontaneous mutations cannot be completely suppressed even in the zero-temperature limit, since they are induced by the driving pulses themselves.
Mutations are rare, as they happen far from resonance.
Because mutations may lead to open-endedness, the model thus hints at a possible connection between dissipative adaptation and open-ended evolution that has not been discussed so far, to the best of my knowledge.

The significance of this paper is mainly to bridge some aspect of artificial life and nonequilibrium thermodynamics.
The quantum mechanical framework has been chosen so as to keep things in terms of basic resources such as atoms, photons, and their interactions.
It turns out that the model may be thought of as illustrative of Walker and Davies viewpoint \cite{algorithmicorigins}, in that genetics is a digital form of information processing (simulated here with quantum bits), to be contrasted with the analogue form of information processing in metabolism (simulated here with continuous pulses).
From a quantum information perspective, this paper can be seen as the quantum thermodynamics of a quantum cloning process \cite{noclo,buzek,gisin,dv2012} driven by single photons.
Even though the words `cloning' and `self-replication' can be used interchangeably here, the option for the latter is motivated by the original goal of mimicking lifelike information processing with an out-of-equilibrium physical system.

\section{Results}

{\bf Quantum artificial organism.}
The quantum organism is defined as a chain of quantum systems, as inspired by the polymeric structure of nucleic acid strands (DNA or RNA).
Each quantum system has three energy levels.
We choose a  lambda configuration, so as to guarantee that the two lowest-energy levels are stable in the zero-temperature limit.
The two lowest-energy levels are labeled here as $\ket{a_n}_1$ and $\ket{b_n}_1$, where $n$ refers to the $n$-th lambda system in the chain, and the index $1$, to the original gene (the original chain).
States $\ket{a_n}_1$ and $\ket{b_n}_1$ play the role of two possible equivalents of nucleotide bases (instead of four, as in DNA or RNA).
The original quantum artificial gene is defined by a (generally aperiodic) sequence of these lowest-energy states (a string), let us say
\beq
\ket{\mathrm{gene}}_1 = \ket{a_1}_1\ket{b_2}_1 \ ... \ \ket{a_N}_1.
\label{gene1}
\eeq
$N$ here is the gene size, also corresponding to the chain size.
Quantum superpositions of bases in $\ket{\mathrm{gene}}_1$ are not considered in this paper, as further explained below.

\

{\bf Self-replication.}
The idea is to look for a dynamical process $U$ (a global organism-plus-environment unitary dynamics), so that
\beq
U\ket{\mathrm{gene}}_1 \ket{\mathrm{bases}}_2 \ket{\mathrm{source}} \ra \ket{\mathrm{gene}}_1 \ket{\mathrm{gene}}_2 \ket{\mathrm{source}'},
\label{sr}
\eeq
where $U$ must be independent of the initial state $\ket{\mathrm{gene}}_1$.
In (\ref{sr}),
$\ket{\mathrm{bases}}_2 = \prod_{n=1}^N \ket{b_n}_2$
is the state representing the available environment bases upon which the organism can act to compose the copied gene, 
$\ket{\mathrm{gene}}_2 $.
States $\ket{b_n}_2$ here are the fundamental states of the lambda systems (therefore, their thermal equilibrium states in the limit of zero temperature).
The source state 
\beq
\ket{\mathrm{source}} = \prod_{n=1}^N \ket{s_n}
\label{source}
\eeq
describes the initial state of the environment degrees of freedom that provide the energy source; by the end of the process, the environment may have been modified to some final state $\ket{\mathrm{source}'} = \prod_n \ket{s_n'}$.
Importantly, the nucleic-acid analogy guides us to look for a process $U =U(\left\{ r_n \right\})$, which depends on the free parameters $\left\{ r_n \right\}$ symbolizing spatial distances between each gene base (the template lambda system) and its corresponding environment base (the environment lambda system which undergoes the copying process).
In other words, the ability of the template lambda system to copy itself shall depend on the distance to the environment base.

\begin{figure}[!htb]
\centering
\includegraphics[width=1.0\linewidth]{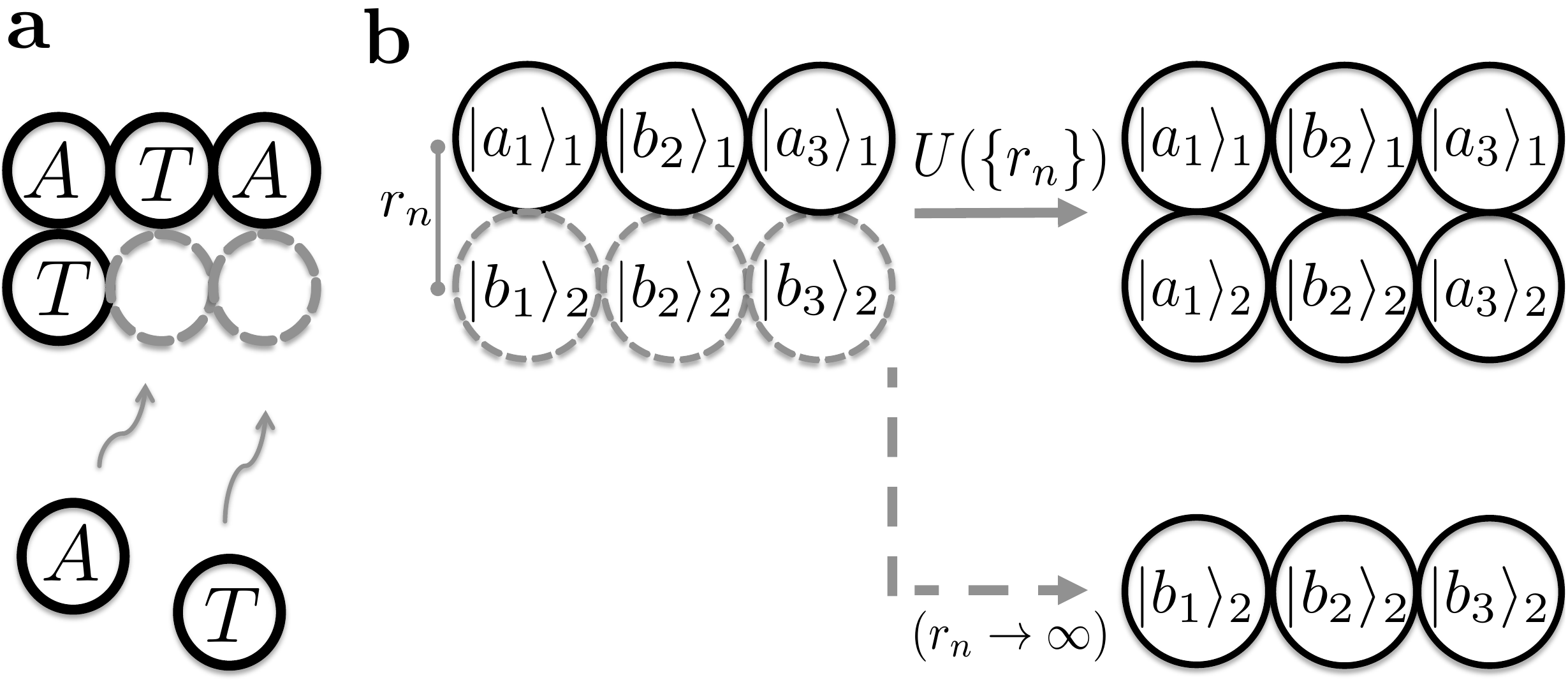} 
\caption{
{\bf Natural versus artificial gene self-replication.}
(a) Simplified picture of a self-replicating biological gene.
A single nucleic acid strand forms the original gene with nucleotide bases $A$, $T$ and $A$ .
The complementary environment bases, $T$, $A$ and $T$, form the copy.
In the process, the bare states (dashed gray circles) undergo a transformation to the filled states (full black circles) dictated by the original gene (curved arrows).
(b) A sequence of states $\ket{a_1}_1\ket{b_2}_1\ket{a_3}_1$ (full black circles) specifies the original quantum artificial gene.
The environment bases are initially set to their ground states, $\ket{b_n}_2$, for $n = 1,2,3$ (dashed gray circles), which can be regarded as bare states.
In the process, the bare states undergo a transformation to a filled state (full arrow), dictated by (and ideally identical to) the original gene state, 
$\ket{a_1}_1\ket{b_2}_1\ket{a_3}_1 
\otimes 
\ket{b_1}_2\ket{b_2}_2\ket{b_3}_2 
\ra 
\ket{a_1}_1\ket{b_2}_1\ket{a_3}_1 
\otimes
\ket{a_1}_2\ket{b_2}_2\ket{a_3}_2$.
This transformation is described by a unitary operator $U(\left\{r_n \right\})$ for the global organism-plus-environment dynamics, including that of the source of energy.
The unitary operator $U(\left\{r_n \right\})$ depends on the free parameter $r_n$, representing the distance between the $n$-th gene base and its corresponding environment base.
If $r_n \ra \infty$, the environment bases are expected to be left unchanged (dashed arrow).
}
\label{fig1}
\end{figure}

From now on, replication is assumed to be modular, that is, the replication of each gene unit ($\ket{ \bullet_n }_1$) is independent of the other units, 
$U(\left\{ r_n \right\}) = \prod_n U_n(r_n)$,
where $[U_n,U_m] = 0$ for all $m,n$.
This assumption is motivated by the modular character of the self-replication of nucleic acids.
A perfect replication transition should now read
\beq
U_n \ket{a_n}_1\ket{b_n}_2\ket{s_n} \ra \ket{a_n}_1 \ket{a_n}_2 \ket{s_n'},
\label{sra}
\eeq
and a perfect dormant transition,
\beq
U_n \ket{b_n}_1\ket{b_n}_2\ket{s_n} \ra \ket{b_n}_1 \ket{b_n}_2 \ket{s_n},
\label{srb}
\eeq
for $n = 1, ... , N$.
To guarantee that a gene base does not affect an infinitely far apart environment base, we look for a unitary that satisfies $U_n(r_n\ra \infty) \ra 1$.
See Fig.\ref{fig1}.

Arbitrary superpositions of $\ket{a_n}_1$ and $\ket{b_n}_1$ cannot be copied with an arbitrarily high fidelity (quality), as states the so called no-cloning theorem \cite{noclo,buzek,gisin,dv2012}.
The theorem assumes, however, that it is perfectly possible in principle to do so for a particular (preferred) orthonormal basis.
This explains our choice in Eq.(\ref{gene1}).
Here, we intend to find the best replicator allowed by nature (by quantum mechanics, in this case), with the goal of verifying whether it does consume a maximal amount of work.
In other words, we look for a model for $U_n$, trying to fulfill the perfect cloning as allowed in principle for the eigenenergies of the lambda systems.


\

{\bf Organism-plus-environment Hamiltonian.}
We assume an autonomous dynamics
\beq
U_n = \exp{\left( -i H t /\hbar \right)},
\eeq
where the time-independent global Hamiltonian $H$, describing the organism plus the environment, is given by
\beq
H = H_S + H_I + H_E.
\label{H}
\eeq
$H$ does not depend on the index $n$, meaning that the dynamics of each base follows the same rules.
The system Hamiltonian, $H_S$, describes the original gene base only.
$H_E$ describes a composite environment,
\beq
H_E = H_B + H_F.
\eeq
$H_B$ is the environment base Hamiltonian, and $H_F$ is the electromagnetic field Hamiltonian.
Correspondingly, the interaction Hamiltonian $H_I$ is also composite,
\beq
H_I = H_{SB} + H_{BF}.
\eeq
$H_{SB}$ describes the interaction of the original gene base with the environment base. 
$H_{BF}$ describes the interaction of the environment base with the electromagnetic field.

The explicit expressions for the Hamiltonians are as follows.
First,
\beq
H_S =  \hbar \omega_b^{(0)} \ket{e_n}_1 \bra{e_n}_1 + \hbar \delta_a^{(0)} \ket{a_n}_1 \bra{a_n}_1,
\eeq
where $\ket{e_n}_1$ is the excited state of the $n$-th lambda system in the original artificial organism.
Analogously, the environment base Hamiltonian reads
\beq
H_B =  \hbar \omega_b^{(0)} \ket{e_n}_2 \bra{e_n}_2 + \hbar \delta_a^{(0)} \ket{a_n}_2 \bra{a_n}_2,
\eeq
showing that all lambda systems have been chosen identical to one another.

The coupling Hamiltonian between the system (the original gene base) and the environment base is
\beq
H_{SB} = -J \ \sigma_{a1}^{(z)} \sigma_{a2}^{(z)},
\eeq
where $\sigma_{ai}^{(z)} = \ket{e_n}_i\bra{e_n}_i - \ket{a_n}_i\bra{a_n}_i$.
$H_{SB}$ allows the gene base state to act as a switch to the dynamics of the environment base.
As will become clearer below, the gene can selectively induce the environment base to become energetically resonant with the electromagnetic field, reminding us of an enzyme-like effect.
For the degree of influence to depend on the distance parameters, we consider $J = J(r_n)$.
Condition $U_n(r_n\ra \infty) = 1$ suggests that $J(r_n\ra \infty) = 0$.
Ising-type couplings such as the one used here find widespread applications in protein physics \cite{pnas87,dv20}.
Also, similar position-dependent dipole-dipole couplings generalizing the van der Waals forces have been proposed as a means for constructing biologically-inspired quantum molecular machines, which actively and autonomously self-protect their quantum data against noise from pre-existing non-engineered environments \cite{tg1,tg2}.

The base-field interaction Hamiltonian is given by a dipolar coupling, in the rotating-wave approximation \cite{cohen, dv21, dv2012},
\beq
H_{BF} = -i\hbar g \sum_\omega (a_\omega \sigma_{a2}^\dagger + b_{\omega} \sigma_{b2}^\dagger - \mbox{H.c.}).
\label{HI}
\eeq
The continuum of frequencies, $\sum_\omega \ra \int d\omega \varrho_\omega \approx \varrho \int d\omega$, gives rise to the dissipation rate $\Gamma = 2\pi g^2 \varrho$, in the Wigner-Weisskopf approximation.
Modes $\left\{ a_\omega \right\}$ and $\left\{ b_\omega \right\}$ represent orthogonal quantized field modes.
The raising operators read $\sigma_{a2}^\dagger = \ket{e_n}_2 \bra{a_n}_2$ and $\sigma_{b2}^\dagger = \ket{e_n}_2 \bra{b_n}_2$.
$\mbox{H.c.}$ is the Hermitian conjugate.
The choice for coupling the field only to the environment base ($\ket{\bullet_n }_2$) is justified by the fact that all lambda systems are identical in this model, hence it should ideally make no difference what base the photon hits each time.
Finally, the field Hamiltonian is
\beq
H_F = \sum_\omega \hbar \omega (a_\omega^\dagger a_\omega + b_\omega^\dagger b_\omega),
\eeq
also considered in the continuum of frequencies limit.

\

{\bf Single-photon pulses as the energy source.}
The source of energy for the self-replication is the initial out-of-equilibrium state of the electromagnetic environment.
To make it the most elementary energy source, we consider in Eq.(\ref{source}) single-photon pulses added to a zero-temperature background \cite{dv21}, so that
\beq
\ket{s_n} = \ket{1^b} \equiv \sum_\omega \phi_\omega^b(0) b_\omega^\dagger \ket{0}.
\label{1b}
\eeq
$\ket{0} = \prod_\omega \ket{0_\omega^a}\ket{0_\omega^b}$ is the vacuum state of all the field modes.
Modes $\left\{ a_\omega \right\}$ are not initially populated, so as to maximize the irreversibility of the self-replication.
The single-photon pulse admits a one-dimensional real-space representation,
\beq
\phi(z,t) \equiv \sum_\omega \phi_\omega^b(t) \exp(i k_\omega z),
\label{realspace}
\eeq
where $k_\omega = \omega/c$, and $c$ is the speed of light.
As before, the sum over modes is to be considered in the continuum of frequencies limit.
The pulse can also be decomposed as
\beq
\phi(z,t) = \phi^{\mathrm{e}}(z,t) \exp{[-i\omega_L (t - z/c)]},
\label{phi0}
\eeq
in terms of its central frequency $\omega_L$ and its envelope function $\phi^{\mathrm{e}}(z,t)$.

To keep the spirit of an autonomous scenario, the photon pulse is assumed to have been spontaneously emitted from an arbitrarily distant source (not considered in the Hamiltonian). This implies an exponential shaped envelope
\beq
\phi^{\mathrm{e}}(z,0) = \mathcal{N} \Theta(-z) \exp{\left[ \Delta z / (2c) \right] }.
\label{spem}
\eeq 
$\Delta$ here is the pulse spectral linewidth.
$\mathcal{N} \equiv \sqrt{2\pi \varrho \Delta}$, a normalization constant.
$\Theta(z)$, the Heaviside step function.
The transition frequency of this hypothetical distant emitter corresponds to $\omega_L$, in Eq.(\ref{phi0}), and its natural linewidth, to $\Delta$.
Most importantly, $\omega_L$ shall be regarded here as a fixed constant, rather than a free-varying parameter.
This is analogous to the idea of a steady peak in the sunlight spectrum.
By contrast, the spectral linewidth $\Delta$ of each photon pulse may vary, so as to mimic a kind of disorder in the natural linewidths in the hypothetical ensemble of distant single-photon emitters.

\

{\bf Global organism-plus-environment dynamics.}
The global dynamics can be described by the state
\beq
\ket{G_k(t)} = \exp(-iHt/\hbar) \ket{G_k(0)},
\label{G}
\eeq
where $H$ is given by Eq.(\ref{H}), and
\beq
\ket{G_k(0)} = \ket{k_n}_1\ket{b_n}_2\ket{1^b}, 
\eeq
for $k=a,b$.

Due to conservation in the number of excitations, the time-dependent state can be written for $k=a$ as
\begin{align}
\ket{G_a(t)} &= \sum_\omega F_\omega(t) \ket{a_n}_1\ket{b_n}_2\ket{1_\omega^b} \nn\\
&+ R_e(t) \ket{a_n}_1\ket{e_n}_2\ket{0} \nn\\
&+ \sum_\omega R_{a \omega}(t) \ket{a_n}_1\ket{a_n}_2\ket{1_\omega^a}.
\label{Ga}
\end{align}
$F_\omega$ describes a failed replication, leaving a photon at mode $b_\omega$.
$R_e$ describes a replication transient excitation, leaving the field in the vacuum state.
$R_{a \omega}$ describes a replication accomplishment, leaving a photon at mode $a_\omega$.

Accordingly, the global state can be written for $k=b$ as
\begin{align}
\ket{G_b(t)} &= \sum_\omega D_\omega(t) \ket{b_n}_1\ket{b_n}_2\ket{1_\omega^b} \nn\\
&+ M_e(t) \ket{b_n}_1\ket{e_n}_2\ket{0} \nn\\
&+ \sum_\omega M_{a \omega}(t) \ket{b_n}_1\ket{a_n}_2\ket{1_\omega^a}.
\label{Gb}
\end{align}
$D_\omega$ describes a dormant state, leaving a photon at mode $b_\omega$.
$M_e$ describes a mutation transient excitation, leaving the field in the vacuum state.
$M_{a \omega}$ describes an (undesirable) mutation accomplishment, leaving a photon at mode $a_\omega$.
The initial state of the field implies that $F_\omega(0) = \phi^{b}_\omega(0)$ or $D_\omega(0) = \phi^{b}_\omega(0)$, as defined in Eq.(\ref{1b}), depending on the initial state of the original gene base.
The equations of motion are shown in the Methods.



\

{\bf Transition probabilities.} The organism replication probability is defined here as
\beq
p_{a_1,b_2 \ra a_1,a_2}(t) \equiv 
\bra{a_n}_1\bra{a_n}_2 \mbox{tr}_F\Big[ \ket{G_a(t)} \bra{G_a(t)} \Big] \ket{a_n}_1\ket{a_n}_2,
\eeq
where $\mbox{tr}_F[\bullet]$ is the partial trace over the field degrees of freedom.
Similarly, the dormant probability is defined as
\beq
p_{b_1,b_2 \ra b_1,b_2}(t) \equiv 
\bra{b_n}_1\bra{b_n}_2 \mbox{tr}_F\Big[ \ket{G_b(t)} \bra{G_b(t)} \Big] \ket{b_n}_1\ket{b_n}_2.
\eeq
By using Eq.(\ref{Ga}), we find that
\beq
p_{a_1,b_2 \ra a_1,a_2}(t) = \sum_\omega |R_{a\omega}(t)|^2.
\eeq
With Eq.(\ref{Gb}), we also find that
\beq
p_{b_1,b_2 \ra b_1,b_2}(t) = \sum_\omega |D_\omega(t)|^2.
\eeq

To obtain explicit expressions, we turn to the one-dimensional real-space representation of the frequency-dependent amplitudes, namely,
$R_a(z,t) \equiv \sum_\omega R_{a \omega}(t) \exp(ik_\omega z)$ 
and 
$D(z,t) \equiv \sum_\omega D_\omega(t) \exp(ik_\omega z)$,
as done in Eq.(\ref{realspace}), and find that
\beq
p_{a_1,b_2 \ra a_1,a_2}(t) = \frac{1}{2\pi \varrho c}\int_{-\infty}^{\infty} |R_a(z,t)|^2 dz,
\label{rep}
\eeq
and 
\beq
p_{b_1,b_2 \ra b_1,b_2}(t) = \frac{1}{2\pi \varrho c}\int_{-\infty}^{\infty} |D(z,t)|^2 dz.
\label{dorm}
\eeq
See the expressions for $R_a(z,t)$ and $D(z,t)$ in the methods.

We now assume the spontaneously emitted photon as described by Eq.(\ref{spem}).
If the initial state of the original base is $\ket{a_n}_1$, we make $\phi(z,0) = F(z,0) = \sum_\omega F_{\omega}(0)\exp(ik_\omega z)$; otherwise (for $\ket{b_n}_1$), we make $\phi(z,0) = D(z,0)$ (defined in the previous paragraph).
We find that (see Methods)
\beq
p_{a_1,b_2 \ra a_1,a_2}(t\ra \infty) = \mathcal{P}(\Delta, \delta_{L-bJ}),
\label{prep}
\eeq
where the detuning is
$\delta_{L-bJ} \equiv \omega_L - \omega_{bJ}$, with reference to the perturbed frequency transition $\omega_{bJ} \equiv \omega_b^{(0)} + J$,
and
\begin{align}
 \mathcal{P}(\Delta, \delta_{L-bJ}) &\equiv 
\frac{\Gamma^2}{\left(\frac{2\Gamma-\Delta}{2}\right)^2+\delta^2_{L-bJ}} \nn \\
&\times \left[ 1 + \frac{\Delta}{2\Gamma} 
- \frac{\Delta(2\Gamma + \Delta)}{\left(\frac{2\Gamma+\Delta}{2}\right)^2+\delta^2_{L-bJ}}\right].
\label{pa1b2a1a2}
\end{align}
We regard only $\Delta$ and $J$ as the truly free parameters in Eq.(\ref{pa1b2a1a2}).
We assume, by contrast, that $\omega_L$ is fixed by the external world, whereas $\omega_b^{(0)}$ and $\Gamma$ are fixed by the internal world.
We are particularly interested in the regimes where $\omega_L - \omega_b^{(0)} \gg \Gamma$ (blue-detuning) and $J>0$, which clarify the picture.

\

\begin{figure}[!htb]
\centering
\includegraphics[width=1.0\linewidth]{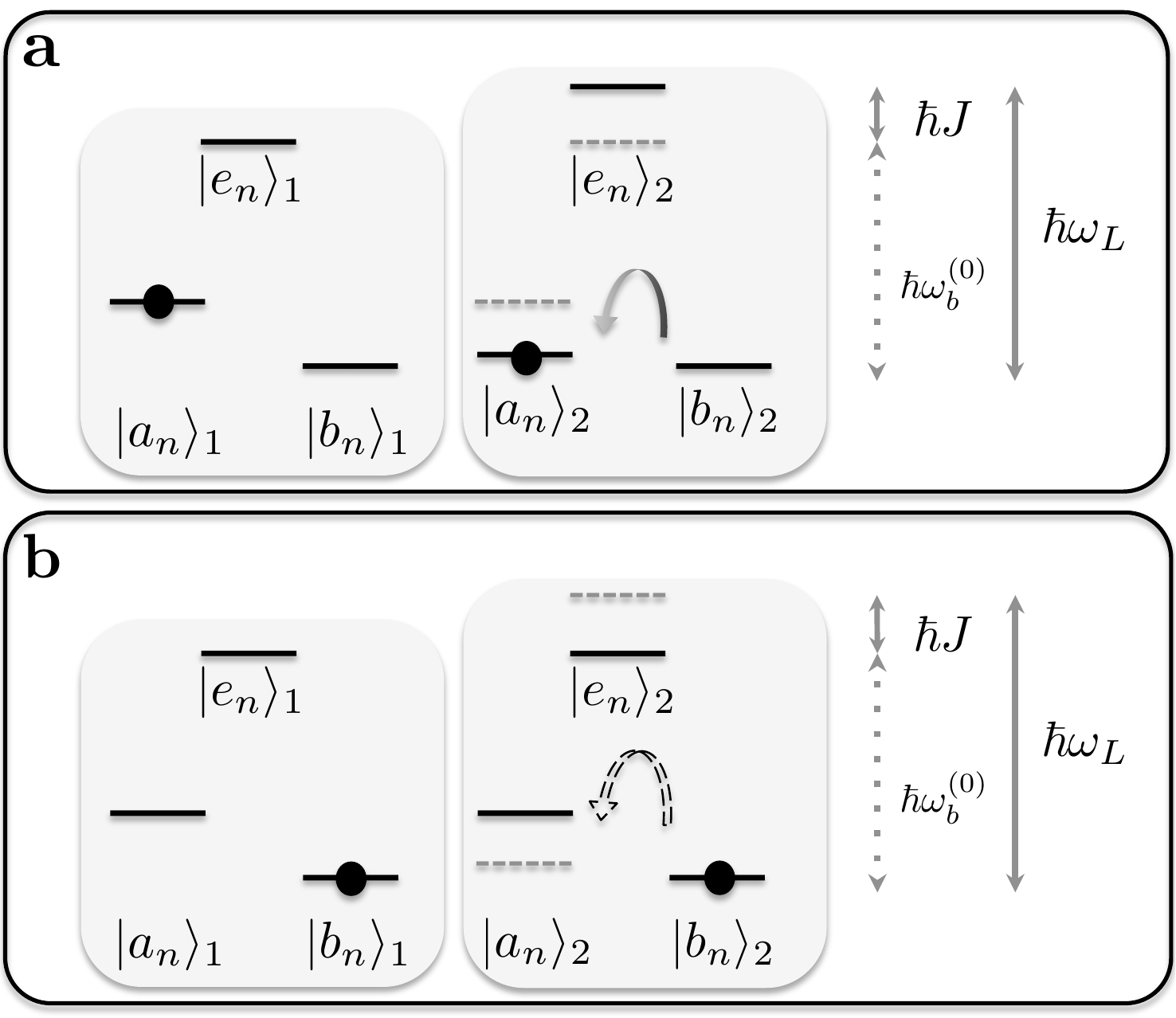} 
\caption{
{\bf Mechanism behind replication and mutation.}
(a) A gene base at $\ket{a_n}_1$ (black dot) induces an energy shift (from the gray to the black horizontal bars) of $\hbar J$ (smaller arrow), building a resonance with the photon, $\hbar \omega_b^{(0)}+\hbar J = \hbar \omega_L$ (longer arrow). $\hbar \omega_b^{(0)}$ is the transition energy of the unperturbed lambda system (dotted arrow).
The environment base thus undergoes a replication transition $\ket{b_n}_2 \ra \ket{a_n}_2$  (full curved arrow).
(b) A gene base at $\ket{b_n}_1$ (black dot) leaves the frequencies unaltered (black horizontal bars), keeping the environment base far from resonance, 
$\omega_b^{(0)} \ll \omega_L$.
However, rare off-resonant photon absorption may induce the mutation transition $\ket{b_n}_2 \ra \ket{a_n}_2$ (dashed curved arrow).
}
\label{fig2}
\end{figure}

{\bf Optimal replication.}
The replication probability at long times, $p_{a_1,b_2 \ra a_1,a_2}(\infty)$, is maximized when the coupling $J=J(r_n)$ induces the resonance condition
\beq
\delta_{L-bJ} \equiv \omega_L - \omega_{bJ} =  \omega_L - (\omega_b^{(0)} + J) = 0.
\eeq
This energy-matching mechanism increasing the likelihood of the process, conditional to the distance $r_n$ between the gene base and the environment base, is reminiscent of an enzyme effect (see Fig.(\ref{fig2})-a).
Perfect replication also requires a monochromatic photon ($\Delta \ll \Gamma$), as we find from Eq.(\ref{pa1b2a1a2}),
\beq
p_{a_1,b_2 \ra a_1,a_2}(\infty)\Big|_{\left\{ \delta_{L-bJ}=0,\ \Delta \ra 0 \right\} } = \mathcal{P}(0,0) = 1.
\eeq
The dormant transition probability is given by
\beq
p_{b_1,b_2 \ra b_1,b_2}(\infty) = 1-\mathcal{P}(\Delta, \delta_{L-b}),
\label{pdorm}
\eeq
where the detuning now refers to the unperturbed frequency, 
\beq
\delta_{L-b} \equiv \omega_L - \omega_b^{(0)},
\eeq
and $\mathcal{P}(\Delta, \delta_{L-b})$ is also defined by Eq.(\ref{pa1b2a1a2}).
The problem of maximizing both the replication and the dormant transition probabilities is discussed in the following.


\

{\bf Mutations.}
Optimizing the replication transition, as described by Eq.(\ref{prep}), and the dormant transition, Eq.(\ref{pdorm}), requires a coupling $J$ that makes the environment base resonant with the photon when the original lambda system is at 
$\ket{a_n}_1$ 
(i.e., $\omega_{bJ} = \omega_L$), 
whereas keeping it far from resonance when the original lambda system is at $\ket{b_n}_1$.
The crucial point here is to notice that, for any finite $J$, an off-resonant photon with respect to the unperturbed frequency ($\delta_{L-b} = J$, since $\omega_{bJ} = \omega_L$) can eventually be absorbed, with the mutation probability (see Fig.(\ref{fig2})-b)
\beq
p_{b_1,b_2 \ra b_1,a_2}(\infty) = \sum_\omega|M_{a\omega}(\infty)|^2 = \mathcal{P}(\Delta, \delta_{L-b}),
\eeq
which vanishes only in the $\delta_{L-b} = J \ra \infty$ limit.
This requires un unphysical photon of infinite frequency to fulfill the perfect replication condition, $\omega_L = \omega_{b}^{(0)}+ J$.
Put differently, equations $p_{a_1,b_2 \ra a_1,a_2}(\infty) = 1$ and $p_{b_1,b_2 \ra b_1, b_2}(\infty) = 1$ cannot be simultaneously satisfied for finite couplings, $J < \infty$.

The finite off-resonant photon absorption probability, leading to $p_{b_1,b_2 \ra b_1, b_2}(\infty) < 1$, is at the root of the unexpected mutations in this model.
It is unexpected because the model was not intentionally designed to present mutations.
On the contrary, we intended to find perfect cloning on a preferred basis, as allowed by the no-cloning theorem \cite{noclo}.
Nevertheless, the idea of a light-induced rare mutation represents here an appealing analogy with natural mutations in biological genes.
According to ref.\cite{nigel}, error in replicating information creates the conditions for the evolvability of life as we know it.
This turns an unfortunate drawback into a powerful resource.

Finally, we note that in the $r_n \ra \infty$ limit (i.e., $J(\infty)=0$), the incoming photon is off-resonant for both initial states of the gene base, $\delta_{L-bJ} = \delta_{L-b} \gg \Gamma$, leading to a predominantly dormant dynamics, $p_{b_1,b_2 \ra b_1,b_2}(\infty) \ra 1$ and $p_{a_1,b_2 \ra a_1,a_2}(\infty) \ra 0$, as expected from the $U(r_n\ra\infty) \ra 1$ condition discussed after Eq.(\ref{srb}).

\

{\bf Dissipative adaptation.}
We finally investigate how dissipative adaptation underlies the self-replication in the present model.
As stated in the introduction, dissipative adaptation is a general thermodynamic mechanism explaining lifelike self-organization in classical far-from-equilibrium systems.
It clarifies how fine-tuned, exceptional behaviours can be fundamentally related with work consumption.
The main picture is that, when a given fluctuating physical system absorbs work from its environment, it can reach exceptional dynamical transformations that are selected by the work source characteristics.
If the excess energy provided by this nonequilibrium work source is dissipated to the environment (in the form of heat), the system can get irreversibly trapped in those rare configurations; in other words, adapted to its environment.
Mathematically, this idea is best described in terms of a fluctuation theorem.
By using Crooks' microscopic reversibility condition for the forward, $p_{i\ra j}(t)$, and backward, $p^*_{j\ra i}(\tau-t)$, classical trajectory probabilities between the initial $i$ and the final $j$ states \cite{crooks}, the dissipative adaptation has been formulated as \cite{naturenano2015,prx2016}
\beq
\frac{p_{i\ra j}(t) }{p_{i\ra k}(t) } = 
e^{-\beta E_{kj}} 
\frac{p^*_{j\ra i}(\tau-t) }{p^*_{k\ra i}(\tau-t) } 
\frac{ \langle e^{-\beta W_{\mathrm{abs}}  } \rangle_{ik}}{\langle e^{-\beta W_{\mathrm{abs}} }  \rangle_{ij} },
\label{cda}
\eeq
where the angle brackets denote a weighted average over all microtrajectories with fixed start $i$ and end $j,k$ points.
$\beta$ is the inverse temperature, $E_{kj} = E_j - E_k$ is the energy difference, and 
$W_{\mathrm{abs}}$ 
is the stochastic nonequilibrium work absorbed from a time-dependent drive.
Equation (\ref{cda}) evidences that a higher work absorption in the $i$ to $j$ transition boosts the probability of state $j$ over the alternative $k$.
Dissipative adaptation can, therefore, describe the self-organization of a physical system enabling it to become apparently well suited to perform some finely-tuned, exceptional task: to seek energy \cite{PRE2015, PRL2017}, to avoid energy \cite{PRL2017, bistableDA,dv21}, or to self-replicate \cite{JCP2013, naturenano2015, prx2016}.

Here, we find quite similar behaviour.
An exceptionally high probability of replication at a zero-temperature environment requires the absorption of a proportionally high amount of average work, as given by
\beq
p_{a_1,b_2 \ra a_1,a_2}(\infty) = \langle W_{\mathrm{abs}} \rangle_{a_1,b_2}/(2\hbar \omega_L).
\label{qdainqsr}
\eeq
$\langle W_{\mathrm{abs}} \rangle_{a_1,b_2}$ is more precisely defined in the following paragraph; its index denotes that the average is calculated with the initial state $\ket{a_n}_1\ket{b_n}_2$ for the matter (and $\ket{1^b}$ for the field, as usual).
Because Eq.(\ref{qdainqsr}) is valid at zero temperature and has been obtained from a fully-quantized model, we can call it a quantum dissipative adaptation \cite{dv21}.
We emphasize that Eq.(\ref{qdainqsr}) is independent of the pulse envelope function (as defined in Eq.(\ref{phi0})), and that it generalizes the result from ref. \cite{dv21}, in being not limited to a resonant photon, and in having been obtained for the dynamics of coupled lambda systems.
\

{\bf Work consumption.}
To obtain our quantum version of the dissipative adaptation, Eq.(\ref{qdainqsr}), we depart from a definition of average incoming work as given by the Heisenberg picture, following ref.\cite{dv21},
\beq
\langle W_{\mathrm{in}} \rangle \equiv \int_0^\infty  \langle \left( \partial_t d(t) \right) E_{\mathrm{in}}(t) \rangle dt,
\label{defwin}
\eeq
where 
$d(t) = U_n^\dagger \ d \ U_n$ is the dipole operator and 
$E_{\mathrm{in}}(t)$ is the incoming-field operator.
Here, 
$d = \sum_i d_{ei} (\sigma_{i2}^\dagger + \mbox{H.c.})$, 
$E_{\mathrm{in}}(t) = \sum_\omega i\epsilon(a_\omega + b_\omega) e^{-i \omega t} + \mbox{H.c.}$, and 
$d_{ei} \epsilon = \hbar g$.
Definition (\ref{defwin}) is very close to our classical notion of work \cite{OL2018,cohen} (to see that, one can think of an initial coherent, or semiclassical, incoming pulse $\ket{\alpha}$, fulfilling $a_\omega\ket{\alpha} = \alpha_\omega \ket{\alpha}$, as established by Glauber \cite{glauber}).
In the rotating-wave approximation, and using the initial global state
$\ket{a_n}_1\ket{b_n}_2\ket{1^b}$, 
we find that (see Methods)
\beq
\langle W_{\mathrm{in}} \rangle_{a_1,b_2} 
= \langle W_{\mathrm{abs}} \rangle_{a_1,b_2} + \langle W_{\mathrm{reac}} \rangle_{a_1,b_2},
\eeq
in terms of the absorptive contribution,
\beq
\langle W_{\mathrm{abs}} \rangle_{a_1,b_2} \equiv 
\hbar \omega_L \int_0^\infty (-2g) \Re[R_e^* e^{- i \delta_a^{ (0) } t} F(-ct,0)] dt,
\label{wabs}
\eeq
and the reactive (dispersive) contribution \cite{OL2018},
\beq
\langle W_{\mathrm{reac}} \rangle_{a_1,b_2} \equiv 
\int_0^\infty (-2\hbar g) \Re[i R_e^* e^{- i \delta_a^{ (0) } t} (\partial_t{F}^e) e^{-i\omega_L t}] dt.
\label{wreac}
\eeq
$\Re[\bullet]$ stands for the real part. 
In Eq.(\ref{wreac}), we have defined $F^e(t)$ such that $F(-ct,0) \equiv F^e(t) \exp(-i\omega_L t)$.
Eqs.(\ref{intre}) and (\ref{modre}) in the Methods, combined with (\ref{wabs}), give us Eq.(\ref{qdainqsr}).

The meaning of the absorptive and reactive work contributions from a single-photon pulse becomes clearer in the monochromatic regime, where an analogy with a classical harmonic oscillator takes place.
The monochromatic regime is set for $\Delta \ll \Gamma$, given the exponential pulse used in Eq.(\ref{prep}). 
We then find that
\beq
r(t) \approx \chi \ f(t),
\label{defchi}
\eeq
where the susceptibility is defined as
\beq
\chi \equiv i \Gamma/(\Gamma - i \delta_{L-bJ}).
\eeq
Here,
$r(t) \equiv \sqrt{\Gamma} R_e(t) e^{i \delta_a^{ (0) } t}$
and
$f(t) \equiv i\sqrt{\Delta}\exp[-(\Delta/2 + i\omega_L)t]$.
This approximately linear dependence is obtained from Eq.(\ref{psimethods}), which depends on the entire history of the photon pulse, revealing the non-Markovianity of the dynamics of the lambda systems \cite{nonmarkov}.
In the monochromatic regime, however, we find that Eq.(\ref{defchi}) holds at times $t\gg \Gamma^{-1}$.
By writing the susceptibility in terms of its real and imaginary parts, 
\beq
\chi = \chi' + i \chi'',
\eeq 
we find that
\beq
\langle W_{\mathrm{abs}} \rangle_{a_1,b_2} \approx 2\hbar \omega_L \ \chi'',
\eeq
and
\beq
\langle W_{\mathrm{reac}} \rangle_{a_1,b_2} \approx \hbar \Delta \ \chi',
\eeq
in close analogy to what has been discussed in refs.\cite{cohen, OL2018}.
Note that the reactive work, 
$\langle W_{\mathrm{reac}} \rangle_{a_1,b_2}$, 
vanishes both at resonance ($\delta_{L-bJ} = 0$), and arbitrarily far from resonance, $\delta_{L-bJ} \ra \infty$ (as expected in dispersive light-matter interactions).
More importantly, it vanishes in the monochromatic limit $\Delta \ra 0$, where the self-replication is optimized.
The absorptive work,
$\langle W_{\mathrm{abs}} \rangle_{a_1,b_2}$,
does not depend on $\Delta \ra 0$, is maximal at resonance, but also vanishes far from resonance.
This tells us that the far-from-resonance (rare) mutations absorb a vanishingly small amount of work.

Analogous results arise from considering a fully classical damped harmonic oscillator with complex position $r_c(t)$, driven by the force $f_c(t)=\sqrt{\Delta}\exp[-(\Delta/2 + i\omega_L) t]$, with a slowly-varying amplitude $\Delta \ll \Gamma$, where $\Gamma$ is the oscillator dissipation rate.
We then have from Newtonian dynamics that $r_c(t) \approx \chi_c f_c(t)$, so the classical work reads 
$W_{\mathrm{cl}} 
\equiv \int_0^\infty 2\Re[f_c^* \partial_t r_c]dt 
= - \int_0^\infty 2\Re[r_c^* \partial_t f_c]dt 
\approx \Delta \chi_c' + 2\omega_L \chi_c''$,
where $\Delta \chi_c'$ is the reactive (dispersive) part, and $2\omega_L \chi_c''$ is the absorptive part (see ref.\cite{cohen}).


\section{Discussion}
In summary, the toy model devised here shows how the nonequilibrium work provided by spontaneously emitted single-photon pulses can fuel the self-replication of an elementary quantum artificial organism formed by a chain of lambda systems.
The guiding intuition was that dissipative adaptation could result in some kind of self-organized process reminding us of an artificial-life code.
Quantum mechanics allowed us to think in terms of the most basic resources in nature, namely, atoms, photons, and their interactions.

Mutations were found to be unavoidable, though rare, due to far-from-resonance photon absorption.
Interestingly, because mutations and self-replication may imply a possible route towards open-ended evolution in biological systems, the model thus alludes to a theoretical link between dissipative adaptation and open-endedness that calls for further investigations.
Finally, it is worth emphasizing that the mutations play a central role in justifying a posteriori why we could assume the existence of an arbitrary state $\ket{\mathrm{gene}}_1$ from the outset.
Put differently, how could state $\ket{a_n}_1$ first have appeared, in an otherwise zero-temperature autonomous universe at thermal equilibrium?
Without mutations, all the bases would perpetually remain in their ground states $\ket{b_n}_i$, implying a completely dormant, rather trivial universe.

As a perspective, we can search for self-organized artificial-life codes with some degree of (quantum or classical) complexity \cite{nicole}.
We can think, for instance, of mimicking the self-organized evolution of an entire artificial genetic code, going beyond the artificial gene we have considered.
To evolve the natural genetic code, biological organisms have taken great advantage from the so called horizontal gene transfer (HGT), according to refs.\cite{goldenfeld,cleland}.
An artificial HGT can be envisioned by letting the free parameters $r_n$ here (representing the distances between pairs of lambda systems) to behave as Brownian particles in a common environment (to be more precise, the center of mass of each lambda system could be considered as a Brownian particle).
Common environments can mediate attractive effective couplings between pairs of Brownian particles (classical and quantum), as shown in refs.\cite{amir,oscar,dv10}.
The size $N$ of each artificial gene would then become a stochastic variable (the artificial gene being slowly split or merged with others in the environment), simply from the environment-induced Brownian movements and effective couplings, thus implementing an artificial HGT.

\section{Methods}
{\bf Solution of the global dynamics.}
The Schr\"odinger equation $i\hbar \partial_t \ket{G_k(t)} = H\ket{G_k(t)}$ (as defined in Eqs.(\ref{H}) and (\ref{G})) leads us to
\begin{align}
\partial_t D_\omega &= -i\omega D_\omega + g M_e, \\
\partial_t M_e &= -i\omega_b^{(0)} M_e - g \sum_\omega (D_\omega + M_{a\omega}), \\
\partial_t M_{a\omega} &= -i(\omega + \delta_a^{(0)}) M_{a\omega} + g M_e,
\end{align}
for $k = a$, 
and
\begin{align}
\partial_t F_\omega &= -i(\omega+\delta_a^{(0)}) F_\omega + g R_e, \\
\partial_t R_e &= -i(\delta_a^{(0)} + \omega_b^{(0)} + J) R_e - g \sum_\omega (F_\omega + R_{a\omega}), \\
\partial_t R_{a\omega} &= -i(\omega + 2\delta_a^{(0)} - J) R_{a\omega} + g R_e,
\end{align}
for $k=b$.

Formally integrating for $D_\omega(t)$, and using $M_{a\omega}(0) = 0$, gives, in the Wigner-Weisskopf approximation, that
\beq
\partial_t M_e = -(\Gamma+i\omega_b^{(0)}) M_e - g D(-ct,0),
\label{meww}
\eeq
where $\Gamma \equiv 2\pi\varrho g^2$ (due to the continuum limit, $\sum_\omega \ra \varrho \int d\omega$), and $D(z,t) \equiv \sum_\omega D_\omega(t)\exp(ik_\omega z)$ (where $\omega = c k_\omega$).
Also, we find that
\begin{align}
D(z,t) &= D(z-ct,0) \nn\\
&+ \sqrt{2\pi \varrho \Gamma} \Theta(z) \Theta(t-z/c) M_e(t-z/c),
\end{align}
and that
\begin{align}
M_a(z,t) &= \sqrt{2\pi \varrho \Gamma} \Theta(z) \Theta(t-z/c) \nn\\
& \times M_e(t-z/c) e^{-i\delta_a^{(0)}z/c}.
\label{mazt}
\end{align}

The results for $R_e(t)$, $F(z,t)$ and $R_a(z,t)$ follow quite similarly,
\beq
\partial_t R_e = -\left( \Gamma+i(\omega_{bJ} +\delta_a^{(0)}) \right) R_e - g F(-ct,0),
\label{reww}
\eeq
where $\omega_{bJ} \equiv \omega_b^{(0)} + J$,
\begin{align}
F(z,t) &= F(z-ct,0) e^{-i\delta_a^{(0)} t}\nn\\
&+ \sqrt{2\pi \varrho \Gamma} \Theta(z) \Theta(t-z/c) R_e(t-z/c) e^{-i\delta_a^{(0)} z/c},
\end{align}
and
\begin{align}
R_a(z,t) &= \sqrt{2\pi \varrho \Gamma} \Theta(z) \Theta(t-z/c) \nn\\
& \times R_e(t-z/c) e^{ -i(2\delta_a^{(0)} - J) z/c}.
\label{eqrazt}
\end{align}

The mutation probability reads
\beq
p_{b_1,b_2 \ra b_1,a_2}(t) = \sum_\omega |M_{a\omega}(t)|^2 
= \frac{1}{2\pi \varrho c}\int_{-\infty}^\infty |M_a(z,t)|^2.
\label{probma}
\eeq
By substituting Eq.(\ref{mazt}) in (\ref{probma}), and changing variables, we find that
\beq
p_{b_1,b_2 \ra b_1,a_2}(t) = \Gamma \int_{0}^{t} |M_e(t')|^2 dt'.
\eeq
Similarly, by substituting Eq.(\ref{eqrazt}) in the replication probability, namely,
\beq
p_{a_1,b_2 \ra a_1,a_2}(t) = \frac{1}{2\pi \varrho c}\int_{-\infty}^{\infty} |R_a(z,t)|^2 dz,
\label{razt}
\eeq
we find that
\beq
p_{a_1,b_2 \ra a_1,a_2}(t) = \Gamma \int_{0}^{t} |R_e(t')|^2 dt'.
\label{intre}
\eeq
The integration above is performed with the solution of $R_e(t)$, that is,
\beq
R_e(t) = - g \int_0^t F(-ct',0) e^{-\left( \Gamma + i(\omega_{bJ}+\delta_a^{(0)}) \right)(t-t')} dt'.
\label{psimethods}
\eeq
Eqs.(\ref{phi0}) and (\ref{spem}) allow $R_e(t)$ to be analytically obtained.
Eq.(\ref{intre}) is a key step for the quantum dissipative adaptation relation, Eq.(\ref{qdainqsr}), as explained below.

\

{\bf Calculating the incoming work.}
Using integration by parts, we can rewrite Eq.(\ref{defwin}) as $\langle W_{\mathrm{in}} \rangle = -\int_0^\infty  \langle d(t) \partial_t E_{\mathrm{in}}(t) \rangle dt$.
In the present model, this results in
\begin{align}
\langle W_{\mathrm{in}}\rangle = -\int_0^\infty dt \ &(i\hbar g) \sum_\omega (-i\omega) \langle \sigma_{a2}^\dagger(t) a_\omega \rangle e^{-i\omega t} \nn\\
&+ (-i\omega) \langle \sigma_{b2}^\dagger(t) b_\omega \rangle e^{-i\omega t} \nn\\
&+ \mbox{c.c.},
\end{align}
where $\mbox{c.c}$ stands for complex conjugate.
Choosing $\ket{1^b}$ as the initial state of the field implies that $\langle \sigma_{a2}^\dagger(t) a_\omega \rangle = 0$.
The non-zero correlation function gives
\begin{align}
\langle \sigma_{b2}^\dagger(t) b_\omega \rangle 
&= \langle U^\dagger \sigma_{b2}^\dagger U  b_\omega \ket{a_n}_1\ket{b_n}_2\ket{1^b} \\
&= \langle U^\dagger \sigma_{b2}^\dagger U  F_\omega(0) \ket{a_n}_1\ket{b_n}_2\ket{0} \\
&= \langle U^\dagger \sigma_{b2}^\dagger e^{-i\delta_a^{(0)} t}  F_\omega(0) \ket{a_n}_1\ket{b_n}_2\ket{0} \\
&= \langle U^\dagger e^{-i\delta_a^{(0)} t}  F_\omega(0) \ket{a_n}_1\ket{e_n}_2\ket{0} \\
&= \langle G_a(t)| \left(\ket{a_n}_1\ket{e_n}_2\ket{0}\right) e^{-i\delta_a^{(0)} t}  F_\omega(0) \\
&= R_e^*(t) e^{-i\delta_a^{(0)} t}  F_\omega(0).
\end{align}
Finally,
\beq
\langle W_{\mathrm{in}}\rangle = 
-\hbar g \int_0^\infty R_e^*(t) e^{-i\delta_a^{(0)} t} \left( i \partial_t F(-ct,0) \right) + \mbox{c.c.}.
\label{windetail}
\eeq
We define $F(-ct,0) = F^e(-ct,0)\exp(-i\omega_L t)$, and obtain Eqs.(\ref{wabs}) and (\ref{wreac}).
To find Eq.(\ref{qdainqsr}), we get from Eq.(\ref{reww}) that
\beq
\partial_t |R_e(t)|^2 = -2\Gamma |R_e(t)|^2 - 2 g \Re[R_e^*(t) e^{-i\delta_a^{ (0) } t } F(-ct,0)].
\label{modre}
\eeq
We substitute Eq.(\ref{modre}) in Eq.(\ref{intre}), using that $R_e(0) = R_e(\infty) = 0$.
Finally, we identify the real part in Eq.(\ref{modre}) with that in the absorptive term, Eq.(\ref{wabs}), which follows from Eq.(\ref{windetail}).
This leads to Eq.(\ref{qdainqsr}), independently of the choice for $F(-ct,0)$, that is, the initial photon wavepacket.

\

\section{Data availability}
Data sharing not applicable to this article as no datasets were generated or analysed during the current study.

\

\section{Code availability}
Code availability not applicable to this article as no codes were generated or analysed during the current study.

\

\section{Author contributions}
D. V. did the research, prepared the figures and wrote the paper.

\section{Competing interests}
The author declares no competing interests.

\

\section{acknowledgements}
I thank Thiago Werlang and Thiago Guerreiro for comments on the manuscript.
This work was supported by the Serrapilheira Institute (Grant No. Serra-1912-32056) and 
by the INCT de Informa\c c\~ao Qu\^antica, CNPq INCT-IQ (465469/2014-0), Brazil.


%

%

%

%

%
\end{document}